\documentclass[
aps, %
prl,
preprintnumbers,
twocolumn,
superscriptaddress,
nofootinbib,
floatfix,
10pt
]{revtex4-1}

\usepackage{amsfonts,amssymb,stmaryrd,latexsym,amsmath,braket}
\usepackage{graphicx,subfigure}
\usepackage{times}
\usepackage{slashed}
\usepackage{braket}
\usepackage{verbatim}
\usepackage[utf8]{inputenc}

\usepackage{color}
\usepackage[dvipsnames]{xcolor}
\usepackage{hyperref}
\DeclareRobustCommand{\citeSM}{\cite{SM}\nocite{Wang:2012eof,Angeli:2013epw,CREX:2022kgg,PREX:2021umo,Zalewski:2008is,Zhao:2024gjz,Kortelainen:2014UNEDF2,Bsk16,Gs}}
\usepackage{bm}
\usepackage{soul}
\usepackage{ulem}

\begin{document}

\title{Bayesian Inference of the Landau Parameter $G'_0$ from Joint Gamow-Teller Measurements}

\author{Zidu Lin}
\affiliation{Sino-French Institute of Nuclear Engineering and Technology, Sun Yat-Sen University, Zhuhai 519082, China}
\affiliation{Department of Physics and Astronomy, University of Tennessee Knoxville, Knoxville, TN 37996-1200, USA.}
\affiliation{Department of Physics and Astronomy, University of New Hampshire, Durham, NH 03824, USA.}

\author{Gianluca Col\`o}
\affiliation{Dipartimento di Fisica, Universit\`a
degli Studi di Milano, via Celoria 16, 20133 Milano, Italy,}
\affiliation{INFN, Sezione di Milano, via Celoria 16, 20133 Milano, Italy}

\author{Andrew W. Steiner}
\affiliation{Department of Physics and Astronomy, University of Tennessee Knoxville, Knoxville, TN 37996-1200, USA.}
\affiliation{Physics Division, Oak Ridge National Laboratory,TN 37831, USA.}

\author{Amber Stinson}
\affiliation{Department of Physics and Astronomy, University of Tennessee Knoxville, Knoxville, TN 37996-1200, USA.}

\begin{abstract}

$G'_0$ is the leading-order Landau-Migdal parameter that characterizes the spin-isospin dependent nucleon-nucleon interaction. Consequently, the $G'_0$ is closely related to the Gamow-Teller resonance (GTR), the beta and double-beta decay rates of finite nuclei, the response of hot and dense nucleonic matter that modifies the neutrino-nucleon absorption rates in core-collapse supernovae (CCSNe) and binary neutron star (BNS) mergers, and finally the critical density for pion condensation in neutron stars. In this letter, for the first time, we report the $G'_0$ with quantified uncertainty in the framework of Bayesian inference, using a self-consistent standard Skyrme Random Phase Approximation (RPA) model and joint constraints from experimental GTR measurements on $^{208}\mathrm{Pb}$, $^{132}\mathrm{Sn}$, $^{90}\mathrm{Zr}$. Our extracted $G_0'$ at saturation density is $0.48\pm0.034$, which is close to the prediction of a few existing Skyrme models that consider spin-isospin observables but smaller than the traditional ones extracted from pion-exchange models. The reasons for this deviation, like the value of the nucleon effective mass $\frac{m^*}{m}$ and the density dependence of $G'_0$, are investigated. The density-dependent $G_0'$ values extracted in this work may guide the construction of new energy density functionals that aims to self-consistently describe the dense matter properties in the spin-isospin channel. 
\end{abstract}

\maketitle

\paragraph{\itshape Introduction.} 

 Understanding the interactions between nucleons is one of the ultimate goals in nuclear physics. The isospin-dependent part of the nuclear force is important for the structure of neutron-rich nuclei, the properties of isospin-asymmetric and pure neutron matter, the evolution and structure of neutron stars. The spin-isospin part of the nuclear force, on the other hand, is crucial for understanding the Gamow-Teller resonance (GTR) in nuclei~\cite{Osterfeld:1991ii,Suzuki:2012ds,PhysRevLett.98.082501} and other charge-exchange (CE) nuclear reactions \cite{Zegers2023}. For example, neutrino interactions with dense matter in core-collapse supernovae and neutron star mergers~\cite{Reddy:1997yr,Burrows:1998cg,PhysRevC.95.025801}, beta-decay and neutrinoless double beta-decay rates in finite nuclei~\cite{PhysRevC.37.731,PhysRevC.55.1532,Suzuki:2012ds,PhysRevLett.114.142501}, and pion condensation in dense matter are all related to the spin-isospin part of the nuclear force~\cite{Migdal:1990vm,Baym:1974vzp}.
 The spin-isospin-dependent nuclear force is characterized by the density-dependent Landau parameter $G'_0 (\rho)$~\cite{Landau:1956zuh}, and is believed to be repulsive at densities relevant to nuclei and nuclear matter in the inner crust/outer core region of compact stars. This argument is supported by the experimental evidence that the residual interaction pushes up in energy the neutron-proton spin-flip transitions: {for instance, the $\nu h^{-1}_{11/2}-\pi h_{9/2}$ or $\nu i^{-1}_{13/2} - \pi i_{11/2}$ transitions that lie at $\approx$ 12 MeV in  
 $^{208}\mathrm{Pb}$ become component of a 
 Gamow-Teller peak at 19.2 MeV.} Consequently, $G_0'(\rho)$ should be positive but its quantitative determination remains challenging if phenomenological models are used, together with either uncontrolled approximations or simplistic uncertainty quantification.

 \begin{figure*}
    \centering
    \includegraphics[width=0.9 \textwidth]{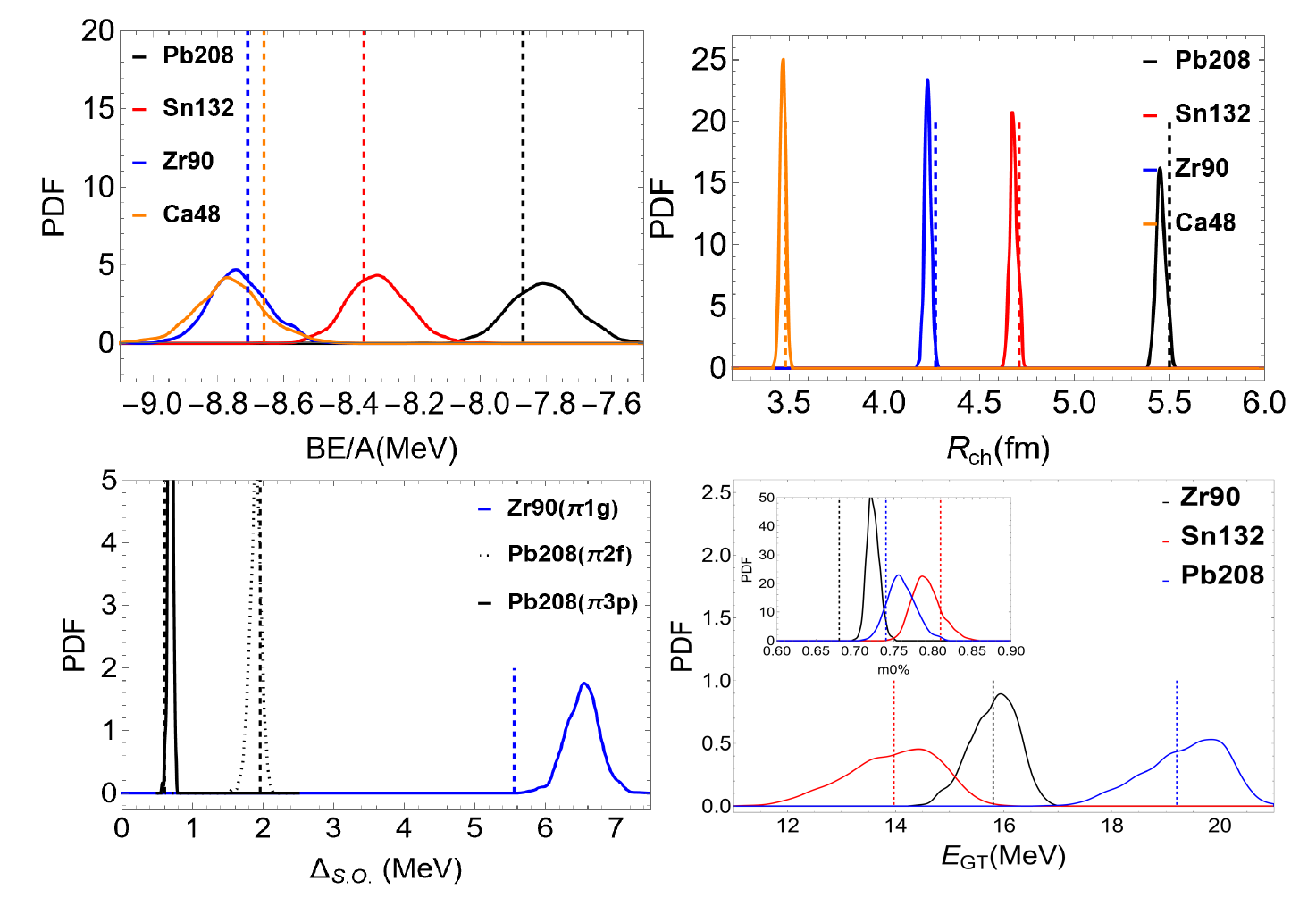}
    \caption{The probability density distribution of nuclear binding energies (upper left), charge radii (upper right), spin-orbit splittings (lower left), $E_\mathrm{GT}$ and $m_0\%$ (lower right),  sampled from GTAll posterior distribution.
    }
\label{fig:baseline}
\end{figure*}

Historically, $G'_0(\rho_0)$ was extracted by fitting it to the peak location of the experimental GTR spectrum $E_\mathrm{GT}$. To relate the Landau parameter with the $E_\mathrm{GT}$, often schematic $\pi$- (or $\pi+\rho$-) exchange models were employed ~\cite{Suzuki:1982usa,Bender:2001up,Towner:1987zz,Osterfeld:1991ii}, assuming 
experimental or Woods-Saxon single-particle states and either a simplified version of linear response theory or a phenomenological RPA. However, the lack of self-consistency  limits the predictive power of these phenomenological models. For example, they cannot predict the GTR features of those nuclei whose single-particle states have not been measured, and assume the value of $G'_0(\rho)$ are the same at all densities. In some of these models, a constant attenuation factor was introduced to estimate the finite-nuclei surface effect \cite{Suzuki:1982usa}. In the following, for simplicity, the $G'_0(\rho_0)$ at nuclear saturation density is denoted at $G'_0$ (unless we discuss explicity its density dependence).  It's also worth mentioning that the extracted dimensionless $g'$ reported in phenomenological $\pi$- (or $\pi+\rho$-) exchange models is not the commonly defined Landau Migdal parameter $G'_0$, but their relationship is $\frac{m}{m^*}G_0^\prime \cdot 150\ {\rm MeV\ fm}^3\ = g^\prime \cdot392\ {\rm MeV\ fm}^3$~\cite{Suzuki:1982usa,Bender:2001up,Towner:1987zz,Osterfeld:1991ii}. Again, without an underlying self-consistent model,
one has to roughly estimate the value of $m^*$ to obtain $G'_0$ given $g'$, introducing additional uncontrolled uncertainties.  

The Skyrme energy density functionals (EDFs) have been proved successful in describing 1) ground state properties of finite nuclei; 2) GTR features of finite nuclei; 3) bulk properties of uniform dense matter in massive stars. Thus, Skyrme EDF is a suitable theoretical framework to self-consistently extract $G'_0$ from the measurements of GTRs. The GTR-informed Skyme EDFs may also provide robust predictions of astrophysical phenomena such as pion condensation and neutrino-dense matter reactions. 

In this letter, we perform the first self-consistent inference of $G'_0$, constrained by joint GTR measurements and without making additional assumptions. Comparing to previous works, the one-to-one relationship between $E_\mathrm{GT}$ and $G'_0$ disappears and $G'_0$ cannot be deterministically fixed due to the large degrees of freedom of the Skyrme EDFs. Consequently, a Bayesian inference like that presented in this letter is necessary to determine the probabilistic distribution of $G'_0$. With the help of Bayesian inference, for the first time, we are able to extract $G'_0$ from multiple GTR experiments with quantified statistical uncertainties. We are also able to systematically analyze the correlations between the predictions of the GTR features and the predictions of ground-state properties of self-consistent Skyrme models \cite{Zlin2025companion}.

\paragraph{\itshape Methods.} 

We first briefly introduce the self-consistent Random Phase Approximation (RPA) model used in this work, which has been widely applied for describing Giant resonances of finite nuclei~\cite{Fracasso:2007fi, PhysRevC.72.064310,Colo2013}. In \cite{Fracasso:2007fi,Colo2013}, more details are provided. The RPA calculations were performed on the basis of a Hartree-Fock (HF) scheme using Skyrme interactions. Given the 
energies $\epsilon$ and wave functions of the HF single-particle states, one can build the QRPA matrix equation
\begin{equation}
    \begin{pmatrix}
        A & B \\
        -B^* & -A^*  
    \end{pmatrix}\begin{pmatrix}
        X^\nu \\
        Y^\nu
    \end{pmatrix}=E_\nu\begin{pmatrix}
        X^\nu \\
        Y^\mu
    \end{pmatrix}~,
\end{equation}
on a particle-hole (p-h) basis. The matrix elements $A$ and $B$ are defined as
\begin{equation}
    A_{mi,nj}=(\epsilon_m-\epsilon_i)\delta_{mn}\delta_{ij}+\langle mj|V_{res}|in\rangle,
\end{equation}
\begin{equation}
    B_{mi,nj}=\langle mn|V_{res}|ij\rangle,
\end{equation}
where occupied (unoccupied) states are denoted as $i, j$ ($m,n$). The $X^\nu$ and $Y^\nu$ are the amplitudes of the eigenvector of the RPA matrix and $E_\nu$ is the corresponding eigenvalue (excitation energy). $V_{res}$ is the residual p-h interaction.

After solving the RPA matrix equation, the 
Gamow-Teller strength function 
is obtained from the eigenstates $|\nu\rangle$, as
\begin{equation}
    S(E)=\sum_n |\langle \nu||\hat{O}_\mathrm{GT}||0\rangle|^2\delta(E-E_\nu),
\end{equation}
where $\hat{O}^\pm_\mathrm{GT}=\sum_{i=1}^A \sigma^i_\pm \tau_\pm^i $. The density-dependent Landau-Migdal parameter $G'_0(\rho)$, is the dominant part in the RPA matrix elements $A$ and $B$~\cite{Fracasso:2007fi}.

\begin{figure*}
	\centering
	\includegraphics[width=0.45\textwidth]{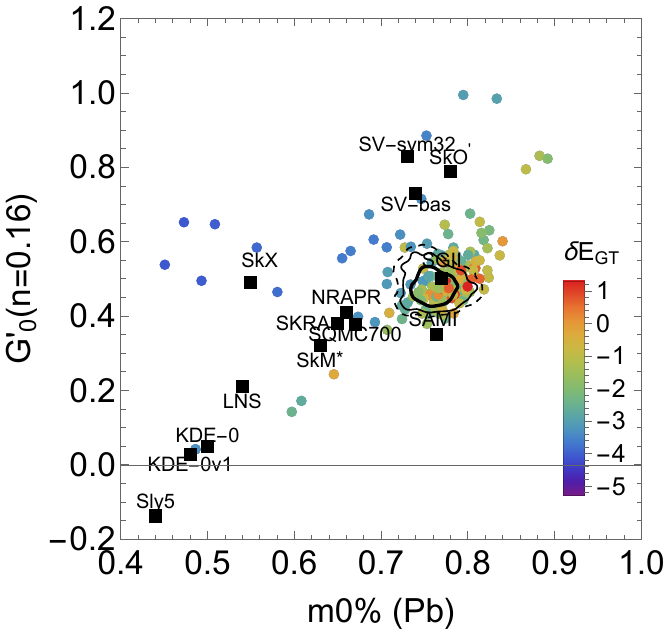}
        \includegraphics[width=0.46\textwidth]{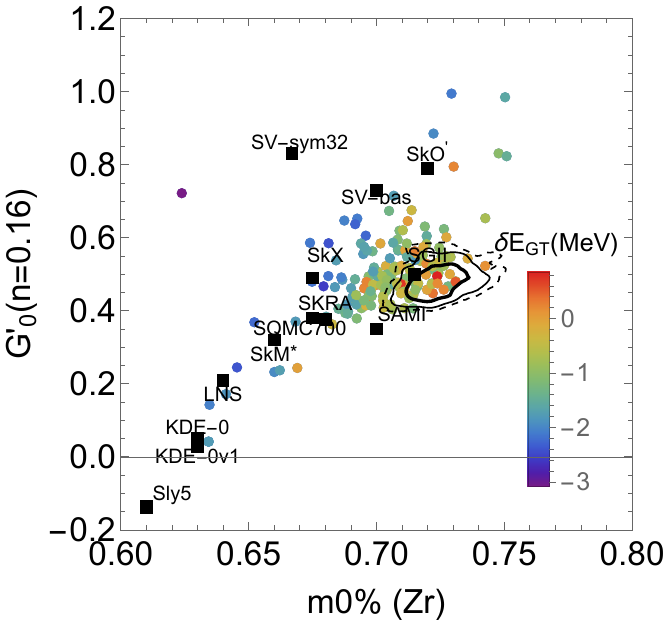}
	\caption{The distribution of $G_0'$, $m_0\%$ and $\delta E_{\mathrm{GT}}$ in GTAll and noGT samples, where the latter is defined as $\delta E_\mathrm{GT}=E^{model}_\mathrm{GT}-E^{exp}_\mathrm{GT}$. The left panel shows the $G'_0$ dependence on $m_0\%$ of $^{208}$Pb. The right panel shows the $G'_0$ dependence on $m_0\%$ of $^{90}$Zr. The colored scattered points are sampled from the Bayesian posterior distribution without Gamow-Teller constraints (noGT). The black solid thick, thin, dashed contours represent the $68\%$, $95\%$, and $99\%$ confidence interval of a posterior distribution including all Gamow-Teller constraints (GTAll). The distribution of $G'_0$ vs $m_0\%$ in $^{132}$Sn behaves similarly as the one in $^{208}$Pb, and it is presented in the supplemental material~\citeSM{} for interested readers. The squares represent 15 Skyrme models (SLy4~\cite{Chabanat:1997un}, SLy5~\cite{Chabanat:1997un}, SV-bas~\cite{PhysRevC.79.034310}, KDE0~\cite{PhysRevC.72.014310}, SAMi~\cite{Roca-Maza:2012dhp}, SkX~\cite{PhysRevC.58.220}, SkM*~\cite{Bartel:1982ed}, SkO'~\cite{PhysRevC.60.014316}, SV-sym32~\cite{PhysRevC.79.034310}, SGII~\cite{vanGiai:1981zz}, KDE0v1~\cite{PhysRevC.72.014310}, LNS~\cite{PhysRevC.73.014313}, NRAPR~\cite{Steiner:2004fi}, SKRA~\cite{2000MPLA...15.1287R}, SQMC700~\cite{Guichon:2006er}). Note that NRAPR predicts $E_\mathrm{GT}$ of $^{90}$Zr equals to 11.3 MeV which is significantly lower than the measurement and is lying beyond the scope of the right panel.  
    } 
\label{fig:GpVsm0percent}	
\end{figure*}

\begin{figure}
	\centering
	\includegraphics[width=0.435\textwidth]{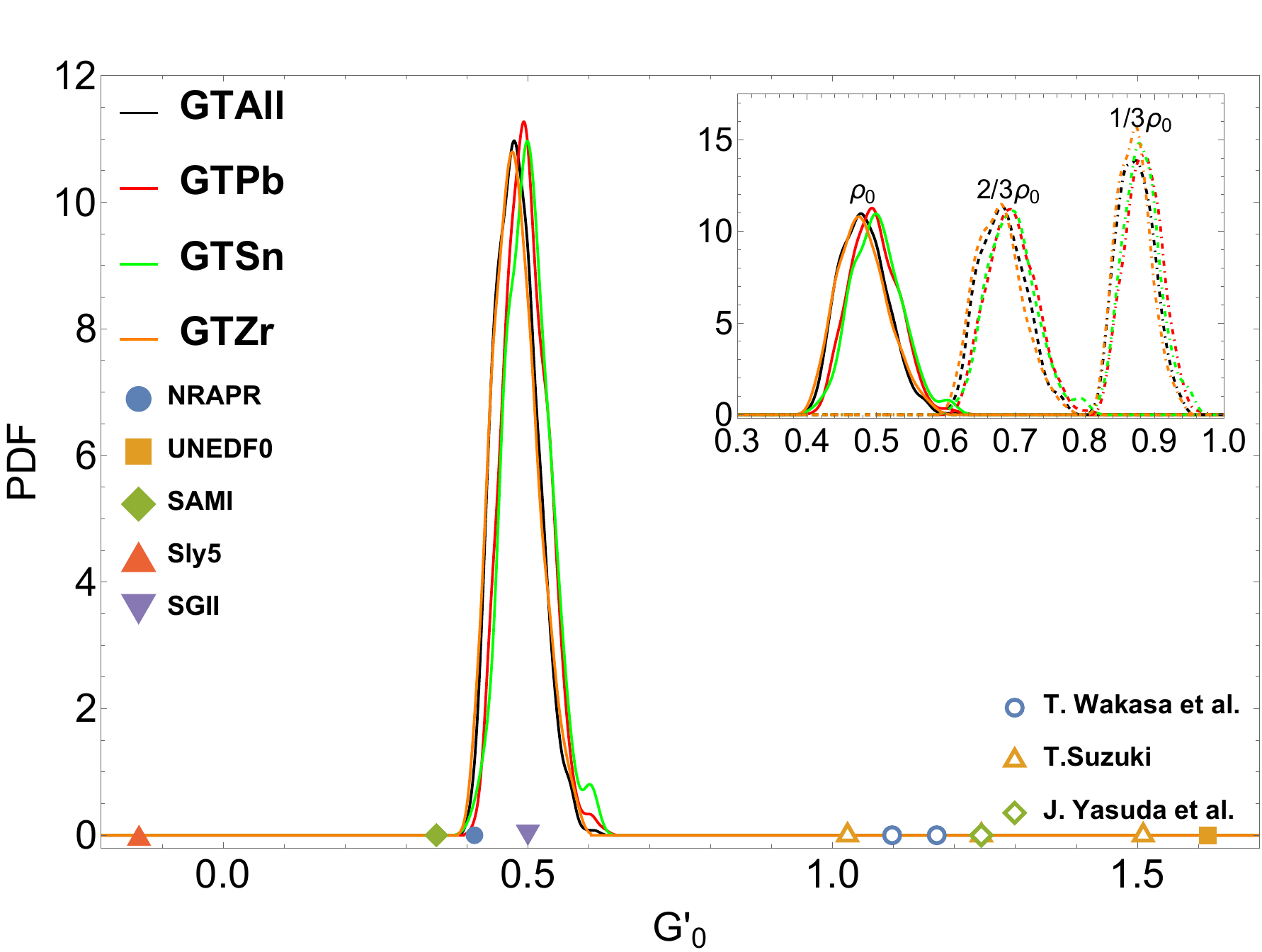}
         \caption{The posterior distribution of $G'_0$. The red, green, orange, and black curves are the posterior of $G'_0(\rho_0)$ extracted from Bayesian inferences with constraints of only $^{208}\mathrm{Pb}$ (GTPb), only $^{132}\mathrm{Sn}$ (GTSn), only $^{90}\mathrm{Zr}$ (GTZr), respectively. The black curve represents the posterior of $G'_0(\rho_0)$ from a Bayesian inference including all the aforementioned GTR measurements (GTAll). The inset shows how the results change if we take a different reference density. The dotted and dot-dashed curves are the distribution of $G'_0$ at $\frac{2}{3}\rho_0$ and $\frac{1}{3}\rho_0$. The filled colored points represent the $G'_0$ of existing  Skyrme models (NRAPR~\cite{Steiner:2004fi}, UNEDF0~\cite{PhysRevC.82.024313}, SAMi~\cite{Roca-Maza:2012dhp}, SLy5~\cite{Chabanat:1997un}, and SGII~\cite{vanGiai:1981zz}) and the hollow colored points represent the extracted $G'_0$ based on phenomenological pion-exchange models (T.Wakasa {\it et al.}~\cite{Wakasa:2004jz,Wakasa:2012zz}, T.Suzuki~\cite{Suzuki:1982usa}, J. Yasuda {\it et al.}~\cite{PhysRevLett.121.132501}). A brief discussion on the difference between the extracted $G'_0$ is given in the supplemental material~\citeSM. 
         } 
\label{fig:GpPosterior}	
\end{figure}

Bayesian inference is a widely used method in nuclear physics and astrophysics to constrain theoretical models given observational data \cite{Rutherford:2024srk,Koehn:2024set,Al-Mamun:2020vzu}. For the first time, the experimental measurements of GTR are used in a state-of-the-art Bayesian inference to constrain Skyrme models. Here, we include the intrinsic scattering (IS) parameters \cite{Nattila:2017wtj,Al-Mamun:2020vzu,Steiner:2017vmg,Nattila:2015jra} that reflect the model systematic uncertainties. The motivation and the technical details of using IS parameters in Bayesian inference is introduced in the supplemental material~\citeSM. The posterior distribution of Skyrme parameterizations are found by:
\begin{equation}\label{eq:postorig}
    P_{\mathrm{post}}(\{p\}) \propto {\cal{L}}(\{p\}) P_{\mathrm{prior}}(\{p\})~,
\end{equation}
where $\{p\}$ is the set of Skyrme parameters and ${\cal{L}}(\{p\})$ is the likelihood constraining the parameter distribution of our model. Given $P_{\mathrm{post}}(\{p\})$, we obtain the posterior distribution of $G'_0$, which is a function of Skyrme parameters~\cite{Fracasso:2007fi}. In this letter, we include constraints of $E_\mathrm{GT}$ and the percentage of Ikeda sum rule $m_0\%$ from $^{208}$Pb, $^{132}$Sn, $^{90}$Zr (denoted as GTAll Bayesian inference) and investigate the dependence of the GT spectral features on $G'_0$ in a self-consistent RPA model. We also perform four additional Bayesian inferences with only GTR constraints from $^{208}$Pb (GTPb), $^{132}$Sn (GTSn), $^{90}$Zr (GTZr), and with no GTR constraints (noGT). Besides the GTR constraints, all Bayesian inferences in this letter are constrained by charge radii and binding energy (from $^{208}$Pb, $^{132}$Sn, $^{90}$Zr and $^{48}$Ca), the neutron skin thickness (in $^{208}$Pb and $^{48}$Ca), and 9 spin-orbit splitting values (in $^{208}$Pb, $^{132}$Sn, $^{48}$Ca). The fraction of GT strength $m\%$ is defined as the ratio of the area covered by the GTR spectrum in the intervals $[15,24]$, $[9,24]$ and $[12,22]$ MeV to the area covered by the spectrum below 31 MeV, for the $^{208}\mathrm{Pb}$, $^{132}\mathrm{Sn}$ and $^{90}\mathrm{Zr}$ GTR spectrum respectively. The setting of Bayesian prior, the constraints involved in the Bayesian likelihood, and the mean of Bayesian posterior distributions are summarized in the supplemental material~\citeSM. A detailed discussion of Bayesian inference is also available in Ref.~\cite{Zlin2025companion}.

\paragraph{\itshape Results.} 

The Bayesian posterior of $G'_0$ is presented and compared with those from previous works. 

We first demonstrate the robustness of the Skyrme models sampled from the GTAll posterior. In Fig.~\ref{fig:baseline}, we present the probability distribution of binding energies per particle ($^{208}$Pb, $^{132}$Sn, $^{90}$Zr, $^{48}$Ca), root mean squared charge radii $R_\mathrm{ch}$ ($^{208}$Pb, $^{132}$Sn, $^{90}$Zr, $^{48}$Ca), spin-orbit splitting $\Delta_{s.o.}$ ($^{208}$Pb, $^{90}$Zr), $E_\mathrm{GT}$ and $m_0\%$ values ($^{208}$Pb, $^{132}$Sn, $^{90}$Zr) of GTAll models. As shown in the upper two panels, the energy per particle and $R_\mathrm{ch}$ of most GTAll models are consistent with experimental measurements within a $\approx 2\%$ precision. In the lower left panel, these models also give a good description of the $\Delta_{s.o.}$ of $^{208}\mathrm{Pb}$ (which is consistent with measurements within an  accuracy of $\approx 0.2~$ MeV). The $\Delta_{s.o.}$  of $^{90}$Zr predicted by GTAll models are slightly larger than the measurements. In the lower right panel, these models give reasonable predictions of the GTR features ($E_\mathrm{GT}$ and $m_0\%$), especially for $^{132}$Sn and $^{208}$Pb. A complete summary of the predicted quantities of the GTAll models and their consistency with experimental measurements is provided in the supplemental material~\citeSM.

In Fig.~\ref{fig:GpVsm0percent},  we illustrate the dependence of $m_0\%$ on $G'_0$ by showing the distributions of $(m_0\%, G'_0)$ sampled from the Bayesian posteriors of both GTAll and noGT. The variance between theoretical and experimental $E_\mathrm{GT}$ is encoded using the colors of the scattering points.  We compare them with existing Skyrme models. The correlation between $G'_0$ and $m_0\%$ is observed both in existing Skyrme models (black squares) and in the noGT posterior (colored scattering dots). The posterior (black contours) of GTAll shows the pinned-down area favored by the GTR measurements. We observe that the correlation between $G'_0$ and $m_0\%$ is obvious in all three cases ($\mathrm{^{208}Pb}$, $\mathrm{^{132}Sn}$ and $\mathrm{^{90}Zr}$), and the correlation strength is slightly higher in $\mathrm{^{90}Zr}$. Another interesting feature is that Skyrme models that predict similar $E_\mathrm{GT}$ values can have very different $m_0\%$ values and vise versa. These degeneracies necessitate the use of both GTR features to precisely constrain $G'_0$, as manifested by the location of GTAll posterior in this figure. In addition, most Skyrme models sampled from the noGT posterior tend to have $E_\mathrm{GT}$ smaller than the experimental measurements. This figure also shows that many successful Skyrme models (black squares) giving good ground state properties~\cite{PhysRevC.85.035201} fail to reproduce the GTR features. This is not surprising, in keeping with the fact that only even-even nuclei are usually in the fitting protocol. Only a few (e.g. SGII, SAMi) predict features close to the posterior constrained by joint GTR experiments. The correlation between $G'_0$ and $E_\mathrm{GT}$ is not strong based on the Bayesian inference, which is also observed in Ref.~\cite{PhysRevC.85.034314}. We observe that in modern self-consistent energy density functionals, $E_\mathrm{GT}$ is not only sensitive to the strength of $G'_0$, but is also influenced by, for example, the single particle spin-orbit splitting values and the effective mass. The uncertainties of these quantities smear the relationship between $E_\mathrm{GT}$ and $m_0\%$. We present 
the afore-mentioned correlations sampled from Bayesian inference in the supplemental material~\citeSM{} to readers interested in this smearing effect. 

Finally, in Fig. \ref{fig:GpPosterior}, based on the GTAll posterior of $G'_0$ (the black solid curve), we conclude that $G'_0=0.48\pm0.034$ at saturation density. The posteriors of $G'_0$ of GTPb (red dashed), GTSn (green dashed), and GTZr (orange dashed), largely overlaps with the GTAll. On the other hand, extracted $G_0^\prime$ from schematic pion-exchange models are not fully consistent with each other \cite{Suzuki:1982usa,Wakasa:2004jz,Wakasa:2012zz,PhysRevLett.121.132501}, as shown by the hollow colored points in this figure. They are systematically larger than the $G'_0$ extracted from self-consistent Bayesian inference. One should note that the $m^*$ inferred from the self-consistent Bayesian inference is smaller than the one usually assumed in schematic models \cite{Suzuki:1982usa}, and a smaller $m^*$ induces a smaller $G'_0$. One should also note that the $G'_0$ extracted from GTR measurements based on schematic models are assumed to be constant at all densities. However, in modern energy density functionals, the $G'_0(\rho)$ is density dependent and the values of $G'_0(\rho)$ at sub-saturation densities are larger than those at saturation density ($G'_0$ can be $\approx0.7$ at $2/3\rho_0$ and $\approx0.9$ at $1/3\rho_0$). Since the GT transition is a surface process, the extracted $G'_0$ from previous works may be an ``effective" constant representing the strength of $G'_0(\rho)$ at sub-saturation densities; from this viewpoint, the tension between extracted values from previous works and this letter is greatly reduced. In the supplemental material~\citeSM{} the posterior distribution of $m^*$, and an example that can shed more light on the difference between schematic models and self-consistent models,  are presented.

\paragraph{Summary.}  

The nonrelativistic Skyrme EDF is one of the few models that can self-consistently describe both the ground state properties and the charge-exchange resonances of finite nuclei. Within the Skyrme framework, systematic calculations of the GT strength and Bayesian inferences are possible while this would be otherwise too challenging.

For the first time, we use Bayesian inference to obtain the probability distribution of $G'_0$. The GTR spectra are calculated in self-consistent Skyrme RPA models and the model parameters are jointly constrained by GTR experiments. Besides GTR, the ground-state properties and spin-orbit splitting values of several representative closed-shell nuclei are also employed to constrain the models. From the posterior distribution, the value of $G'_0$ with quantified uncertainty is $0.48\pm0.034$. This value is close to the prediction of a few set of existing Skyrme models and is smaller than the $G'_0$ values reported in previous works based on pion exchange models. 

We find that for most standard Skyrme EDFs, it is not necessary to have a $G'_0$ close to the ones extracted from phenomenological  models~\cite{Suzuki:1982usa,Wakasa:2004jz,Wakasa:2012zz,PhysRevLett.121.132501} to reproduce the experimental GTR spectrum features. Actually, in standard Skyrme EDFs, models with $G'_0$ obviously larger than 0.5 may produce an overestimate of $m_0\%$.
We should consider that in previous works a few critical assumptions were made as we have explained, and empirical single-particle states are generally input, whereas ours is a self-consistent model. The effective nucleon mass is certainly different in the two cases. 
This warns against the use of hybrid Hartree Fock + RPA models (which use Skyrme parameterization to calculate the mean field and empirically fitted values of $G'_0$ to calculate the RPA residual force ~\cite{PhysRevC.59.2888}). 

The results in this work may guide the construction of Skyrme EDFs aiming to robustly describe collective excitations of dense nuclear matter and nuclei in the spin-isospin channel. The GTAll models (available in \cite{zlin2025zenodo}) may be used to predict the GTR spectrum of other nuclei and the neutrino-nucleus interactions. They can also be used as training data for constructing emulators of future GTR calculations, and
for further searches. The extracted $G'_0$ with quantified uncertainties has been used by us for constraining the calculations of charged current neutrino-nucleon interactions in the dense nucleonic matter of CCSNe and BNS mergers \cite{Zlin2025companion}. 

Future improvements along the line of this work may involve: increasing the accuracy of the description of single-particle states in Skyrme models, investigating the Bayesian inference of $G'_0$ with extended Skyrme models~\cite{Chamel:2009yx,Goriely:2010bm} or self-consistent relativistic Hartree/Hartree-Fock models~\cite{Paar:2004re,Niksic:2004az,Liang:2008tv}, 
and inventing fast and accurate emulators of the GTR spectral features given a set of Skyrme parameters \cite{Jin:2025aam}. This latter step may pave the way towards the inference of Skyrme parameters that can be used beyond RPA (e.g., second RPA, or particle-vibration coupling (PVC) models).
In fact, we emphasize that the present value of $G'_0$ is extracted within the framework of Skyrme EDFs and RPA models, and this extracted value may be different based on different many-body methods. The model dependence of the $G'_0$ extraction is left for future studies.

\subsection*{Acknowledgements}

ZL and AWS were supported by NSF PHY 21-16686. AWS was also supported by the Department of Energy Office of Nuclear Physics. This work used Bridges2 at University of Pittsburgh through allocation PHY230061 from the Advanced Cyberinfrastructure Coordination Ecosystem: Services $\&$ Support (ACCESS) program, which is supported by U.S. National Science Foundation grants $\#2138259$, $\#2138286$, $\#2138307$, $\#2137603$, and $\#2138296$.

\bibliography{References}
\bibliographystyle{apsrev} 

\end{document}



\section*{Supplemental Material for ``Bayesian Inference of the Landau Parameter $G^{\prime}_0$
from Joint Gamow-Teller Measurements''}
\addtocounter{section}{10}

\author{Zidu Lin}
\author{Gianluca Col\`o}
\author{Andrew W. Steiner}
\author{Amber Stinson}



\subsection{Definition of Landau parameters} 

We assume we have a spin-isospin zero-range interaction in the following form:
\begin{equation}
V_{\sigma\sigma\tau\tau} = V\delta(\vec r_1-\vec r_2)\vec\sigma_1\cdot\vec\sigma_2\vec\tau_1\cdot\vec\tau_2.
\end{equation}

The Skyrme p-h interaction that we employ in this work has the same form, as detailed e.g. in Appendix F of~\cite{Bender:2001up}. To extract the Landau parameters $G_\ell^\prime$, one writes the matrix elements of the interaction between plane waves, $\langle \vec k_1 \vec k_2 \vert V_{\sigma\sigma\tau\tau} \vert \vec k_1 \vec k_2 \rangle$. If we restrict to $k_1=k_2=k_F$, these matrix elements can only depend on the angle $\theta$ between $\vec k_1$ and $\vec k_2$, and are conventionally written, in the Landau-Migdal theory, as
\begin{equation}
\langle \vec k_1 \vec k_2 \vert V_{\sigma\sigma\tau\tau} \vert \vec k_1 \vec k_2 \rangle = N_0^{-1} \sum_\ell G_\ell^\prime P_\ell(cos\theta).
\end{equation}

The quantity $N_0$ is the number of states around the Fermi surface per unit volume and unit energy, that is, in
symmetric matter,
\begin{equation}
N_0 = \frac{2k_F m^*}{\pi^2\hbar^2}.
\end{equation}
Then
\begin{equation}
N_0^{-1} = \frac{\pi^2\hbar^2}{2k_F m^*} \approx 150\ {\rm MeV\cdot fm^3}\ \frac{m}{m^*}.
\end{equation}
As mentioned in various papers (Ref.~\cite{Towner:1987zz} and p. 527 of ~\cite{Osterfeld:1991ii}), there are different conventions (like e.g. using the density of states for a Fermi gas with two particles per momentum state: this would produce a value of $N_0^{-1}$ that is a factor of two times larger, around $\approx 300\ {\rm MeV\cdot fm^3}$. The J\"ulich group has used this convention.)

The groups working with a $\pi+\rho$-residual p-h force use a different convention for the Landau parameters because the quantity $N_0^{-1}G_0^\prime$ is replaced by $\frac{f^2_\pi}{m^2_\pi}g^\prime$. 
We follow Ref.~\cite{Towner:1987zz} in using a lowercase letter to indicate the Landau parameter written with this convention. The key quantity is
\begin{equation}
\frac{f_\pi^2}{m_\pi^2} \approx 392\ {\rm MeV\ fm}^3.
\end{equation}

In ~\cite{Towner:1987zz}, the notation $4\pi f_\pi^2$ is used instead of $f_\pi^2$.

Then, the equivalence with the Landau parameter in the previous convention (uppercase letter) reads
\begin{equation}
N_0^{-1}G_0^\prime = 
G_0^\prime \ 150\ {\rm MeV\ fm}^3\ \frac{m}{m^*} = 
\frac{f^2_\pi}{m^2_\pi}g_0^\prime = 
g_0^\prime ~ 392\ {\rm MeV\ fm}^3.
\end{equation}
This is the same as Eq. (29) of ~\cite{Suzuki:2012ds}, where only $\hbar\equiv 1$.

\begin{table}[htb]
\caption{Prior distribution of Skyrme parameters. 
\label{tab:prior}}
\centering
\begin{ruledtabular}
\begin{tabular}{lll}
  parameter & prior \\
\hline
$t_0$ [MeV$\cdot\mathrm{fm}^3$]  & [-3703.94, -751.214]  \\
$t_1$ [MeV$\cdot\mathrm{fm}^5$]  & [38.81, 710.82]     \\
$t_2$[MeV$\cdot\mathrm{fm}^5$] & [-2276.1, 2913.02] \\
$t_3$ [MeV$\cdot\mathrm{fm}^{3+3\alpha}$] & [6632, 20076.5]\\
$\alpha$ & [-0.063, 0.559] \\
$x_0$ & [-1.81, 2.15] \\
$x_1$ & [-7.53, 3.77] \\
$x_2$ & [-49.90, 91.93] \\
$x_3$ & [-3.41, 3.73] \\
$b_4$ [$\mathrm{fm}^4$] &[-0.36, 0.72] \\
$b'_4$ [$\mathrm{fm}^4$] &[-0.36, 0.72] \\
\end{tabular}
\end{ruledtabular}
\end{table}

\begin{table*}[htb]
    \centering
    \begin{ruledtabular}{}
\begin{tabular}{llcccc}
 & Property& Experiment  & Posterior\\
\hline
&$R_{ch} ^{^{48}Ca}$ [fm] & 3.48 (0.069)  & 3.47 (0.015)\\
&$R_{ch} ^{^{90}Zr}$ [fm] & 4.27 (0.085) & 4.23 (0.017)\\
&$R_{ch} ^{^{132}Sn}$ [fm] & 4.71 (0.09) & 4.68 (0.020)\\
&$R_{ch} ^{^{208}Pb}$ [fm] & 5.50 (0.11) & 5.46 (0.025)\\
Bulk properties&BE/A$ ^{^{48}Ca}$ [MeV] & 8.67 (0.17) & 8.76 (0.10)\\
of nuclei ~\cite{Wang:2012eof,Angeli:2013epw} &BE/A$ ^{^{90}Zr}$ [MeV] & 8.71 (0.17)  & 8.73 (0.088) \\
&BE/A$ ^{^{132}Sn}$ [MeV] & 8.355 (0.17)  & 8.31 (0.091)\\
&BE/A$ ^{^{208}Pb}$ [MeV] & 7.87 (0.16)  & 7.80 (0.10)\\
\hline
CREX &$F_\mathrm{ch} ^{^{48}Ca}-F_\mathrm{w} ^{^{48}Ca}$ [] & 0.0277 (0.0055) & 0.039 (0.0028)\\
PREX ~\cite{CREX:2022kgg,PREX:2021umo} &$F_\mathrm{ch} ^{^{208}Pb}-F_\mathrm{w} ^{^{208}Pb}$ [] & 0.041 (0.013) & 0.0229 (0.0038)\\
\hline
 &$\nu 1f^{-1}_{7/2}$-$\nu 1f^{-1}_{5/2}$ ($^{48}\mathrm{Ca}$) [MeV]& 8.41 (0.67) & 6.14 (0.29)\\
 &$\pi 1f_{7/2}$-$\pi 1f^{-1}_{5/2}$ ($^{48}\mathrm{Ca}$) [MeV]& 4.92 (0.39) & 6.70 (0.25)\\
 &$\nu 1g^{-1}_{9/2}$-$\nu 1g^{-1}_{7/2}$ ($^{90}\mathrm{Zr}$) [MeV]& 7.07 (0.57) & 6.01 (0.26)\\
 Spin-orbit&$\pi 1g_{9/2}$-$\pi 1g^{-1}_{7/2}$ ($^{90}\mathrm{Zr}$) [MeV]& 5.56 (0.44) & 6.52 (0.24) \\
 splitting &$\nu 1i^{-1}_{13/2}$-$\nu 1i_{11/2}$ ($^{208}\mathrm{Pb}$) [MeV]& 5.91 (0.42) & 5.82 (0.28) \\
 constraints ~\cite{Zalewski:2008is} &$\nu 3p^{-1}_{3/2}$-$\nu 3p^{-1}_{1/2}$ ($^{208}\mathrm{Pb}$) [MeV]& 0.90 (0.052) & 0.89 (0.042) \\
 &$\nu 2f^{-1}_{7/2}$-$\nu 2f^{-1}_{5/2}$ ($^{208}\mathrm{Pb}$) [MeV]& 2.06 (0.12) & 2.36 (0.11) \\
 &$\pi 2f_{7/2}$-$\pi 2f_{5/2}$ ($^{208}\mathrm{Pb}$) [MeV]& 1.96 (0.053) & 1.92 (0.082)\\
 &$\pi 3p_{3/2}$-$\pi 3p_{1/2}$ ($^{208}\mathrm{Pb}$) [MeV]& 0.607 (0.21) & 0.69 (0.031)\\
 \hline
 & $E_\mathrm{GT}$ ($^{90}\mathrm{Zr}$) & 15.8 (0.32)  & 15.82 (0.43) \\
 & $E_\mathrm{GT}$ ($^{132}\mathrm{Sn}$) & 13.97 (0.28) & 13.97 (0.83)\\
 Gamow-Teller & $E_\mathrm{GT}$ ($^{208}\mathrm{Pb}$) & 19.2 (0.38) & 19.36 (0.74) \\
 constraints& $m_0\%$ ($^{90}\mathrm{Zr}$) & $68\%$ ($1.4\%$) & $72.0\%$ ($0.78\%$)\\
 & $m_0\%$ ($^{132}\mathrm{Sn}$) & $81.0\%$ ($16\%$) & $79.23\%$ ($1.8\%$)\\
 & $m_0\%$ ($^{208}\mathrm{Pb}$) & $74\%$ ($1.48\%$) & $76.0\%$ ($1.7\%$)\\
\end{tabular}
    \end{ruledtabular}
    \caption{The full list of experimental constraints applied in the Bayesian inference of this work. The values inside the brackets are the experimental uncertainties. The fraction of GT strength $m0\%$ is
    defined as the ratio of the area covered by the GTR spectrum in the intervals $[15,24]$, $[9,24]$ and $[12,22]$ MeV to the area covered by the spectrum below 31 MeV, for the $^{208}\mathrm{Pb}$, $^{132}\mathrm{Sn}$ and $^{90}\mathrm{Zr}$ GTR spectrum respectively. The relative uncertainty of accurately measured values (the binding energy, the charge radii and the peak location of GTR of finite nuclei) are artificially chosen to be $2\%$ of the experimental values to ensure the feasibility of Monte Carlo simulations. 
    }
    \label{tab:Bayesian constraint full}
\end{table*}
\subsection{The systematic model uncertainty in Bayesian inference }

Although highly desirable, there is no ``perfect'' nuclear model that can simultaneously describe both the properties of finite nuclei (including the ground state and the exited states) and infinite dense matter properties. The intrinsic scattering (IS) parameters representing the ``unknown systematic uncertainties'' have been widely used in Bayesian studies~\cite{Al-Mamun:2020vzu,Steiner:2017vmg,Nattila:2015jra,Nattila:2017wtj}. 
We explain both the motivation and the physical meaning of these parameters in what follows.

The IS parameters aim to quantify the uncertainty due to our ignorance of theoretical models: by increasing the IS parameters, the possible credible regions for the model parameters (the Skyrme model parameters in the present case) also inflate to take into account that the data are not fully described by the model, and the IS parameters will reach equilibrium when the full dataset used in the inference becomes self-consistent. In this sense, the IS parameters also \emph{quantify the tensions among model predictions} since our incomplete knowledge of the model cause tensions among model predictions. In Bayesian studies, the strength distributions of these IS parameters are purely ``data driven", so one does not need to specify the source of ``unknown systematic uncertainties".   

The ``intrinsic scattering (IS) parameters" are applied in the Bayesian inference of this work as additional degrees of freedom. Here, the IS parameters of $\hat{O}_k$ are defined as the logarithm of the ratio of the systematic uncertainty to the mean of experimental measurement:
\begin{equation}\label{eq:IS}
    \mathrm{IS}=\log_{10} \left(\frac{\sigma_\mathrm{k,sys}}{O_{k,\mathrm{mean}}}\right),
\end{equation}
where $O_{k,\mathrm{mean}}$ is the mean of experimental measurement, and $\hat{O}_k$ is the theoretical prediction. In this work, we have six intrinsic scattering parameters, assuming that the model predictions corresponding to the same intrinsic scattering parameter have no severe tensions between each other and have similar model uncertainties. The six parameters represent systematic uncertainty of $m_0\%$ in finite nuclei, the systematic uncertainty of $E_\mathrm{GT}$, the systematic uncertainty of ground state properties of $^{208}\mathrm{Pb}$ and $^{132}\mathrm{Sn}$, the systematic uncertainty of ground state properties of $^{90}\mathrm{Zr}$ and $^{48}\mathrm{Ca}$, the systematic uncertainty of the spin-orbit splittings of $^{208}\mathrm{Pb}$ and $^{132}\mathrm{Sn}$, and the systematic uncertainty of spin-orbit splittings of $^{90}\mathrm{Zr}$ and $^{48}\mathrm{Ca}$. 

	

\begin{figure*}
	\centering
	\includegraphics[width=0.435\textwidth]{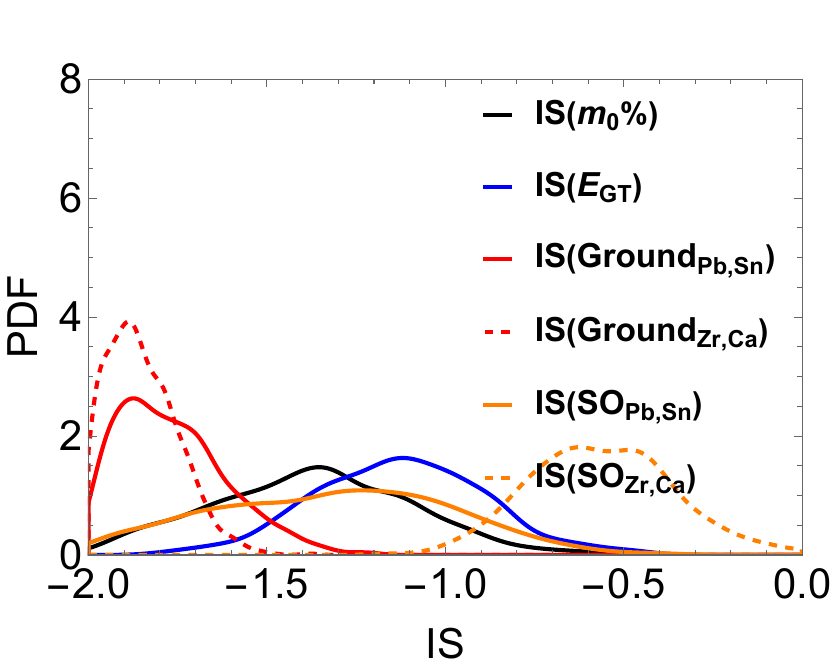}	
	\caption{The posterior of intrinsic scattering parameters in the GTAll Bayesian inference described in the text. 
    } 
\label{fig:ISall}	
\end{figure*}

\begin{figure*}
	\centering
	\includegraphics[width=0.46\textwidth]{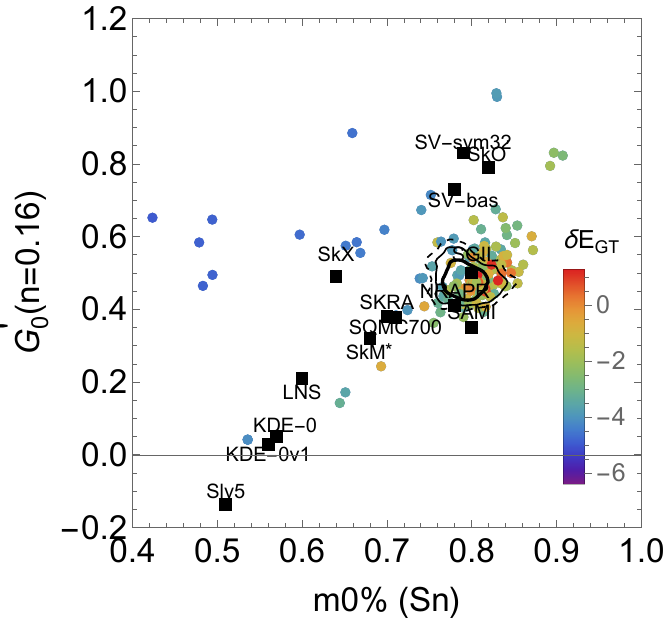}
	\caption{Similar as Fig. 2 of the main manuscript, but for $^{132}$Sn. 
    } 
\label{fig:GpVsm0percentsn}	
\end{figure*}


\subsection{Bayesian prior}
The prior distribution of Skyrme parameters is chosen similarly as in Ref.~\cite{Zhao:2024gjz}.
The ranges of the $t_i$, $x_i$, $\alpha$, $b_4$, and $b_4^{\prime}$ parameters in the Skyrme prior distributions are determined as follows. First, we examined 11 widely used Skyrme models (SGII~\cite{vanGiai:1981zz}, NRAPR~\cite{Steiner:2004fi}, UNEDF0~\cite{PhysRevC.82.024313}, UNEDF2~\cite{Kortelainen:2014UNEDF2}, SLy4~\cite{Chabanat:1997un}, SV-min~\cite{PhysRevC.79.034310}, SkO~\cite{PhysRevC.60.014316}, SkO'~\cite{PhysRevC.60.014316}, BSk16~\cite{Bsk16}, KDE0v~\cite{PhysRevC.72.014310}, and Gs~\cite{Gs}).
We then identified the maximum and minimum $x_i$, $b_4$, and $b'_4$ parameters in these 11 models, denoted as $x_i^{\mathrm{max}}$, $x_i^{\mathrm{min}}$, $b_4^{\mathrm{max}}$, $b_4^{\mathrm{min}}$, $b_4^{\prime \mathrm{max}}$, and $b_4^{\prime \mathrm{min}}$.
To cover a wide range of possible Skyrme parameters, we determine the prior distribution of $t_i$, $x_i$, $\alpha$, $b_4$, and $b'_4$ parameters as $[t_i^{\mathrm{min}} - \Delta t_i, t_i^{\mathrm{max}} + \Delta t_i]$, $[x_i^{\mathrm{min}} - \Delta x_i, x_i^{\mathrm{max}} + \Delta x_i]$, $[\alpha^{\mathrm{min}} - \Delta \alpha, \alpha^{\mathrm{max}} + \Delta \alpha]$, $[b_4^{\mathrm{min}} - \Delta b_4, b_4^{\mathrm{max}} + \Delta b_4]$, and $[b_4^{\prime \mathrm{min}} - \Delta b_4^{\prime}, b_4^{\prime \mathrm{max}} + \Delta b_4^{\prime}]$, where $\Delta t_i = t_i^{\mathrm{max}} - t_i^{\mathrm{min}}$, $\Delta x_i = x_i^{\mathrm{max}} - x_i^{\mathrm{min}}$, $\Delta \alpha = \alpha^{\mathrm{max}} - \alpha^{\mathrm{min}}$, $\Delta b_4 = b_4^{\mathrm{max}} - b_4^{\mathrm{min}}$, and $\Delta b_4^{\prime} = b_4^{\prime \mathrm{max}} - b_4^{\prime \mathrm{min}}$. The prior distribution of all the IS parameters are flatly distributed in $[-2, 0]$.

\subsection{Bayesian Constraints and posterior }

In table.~\ref{tab:Bayesian constraint full}, we summarize the experimental constraints in the GTAll Bayesian inference. The relative uncertainties of the experimental charge radii and binding energy are chosen as $2\%$ to ensure the feasibility of Monte Carlo simulations. The mean values and uncertainty of the spin-orbit splitting are estimated based on the data summarized in ~\cite{Zalewski:2008is}. The averaged values of the GTAll posterior are listed in the last column of the table. 

In Fig.~\ref{fig:ISall}, we present the posterior distribution of the intrinsic parameters. We observe that the IS value of ground-state properties remains small, while the IS value of SO splitting (especially for those of light nuclei) becomes large, which suggests a potential tension between the fitting of SO splitting properties and the fitting of other finite nuclei properties within the Skyrme EDF anstaz.

In Fig.\ref{fig:GpVsm0percentsn}, we present the posterior of $(m_0\%, G'_0)$ similar as Fig. 2, but for $^{132}$Sn.

\begin{figure*}
	\centering
	\includegraphics[width=0.35\textwidth]{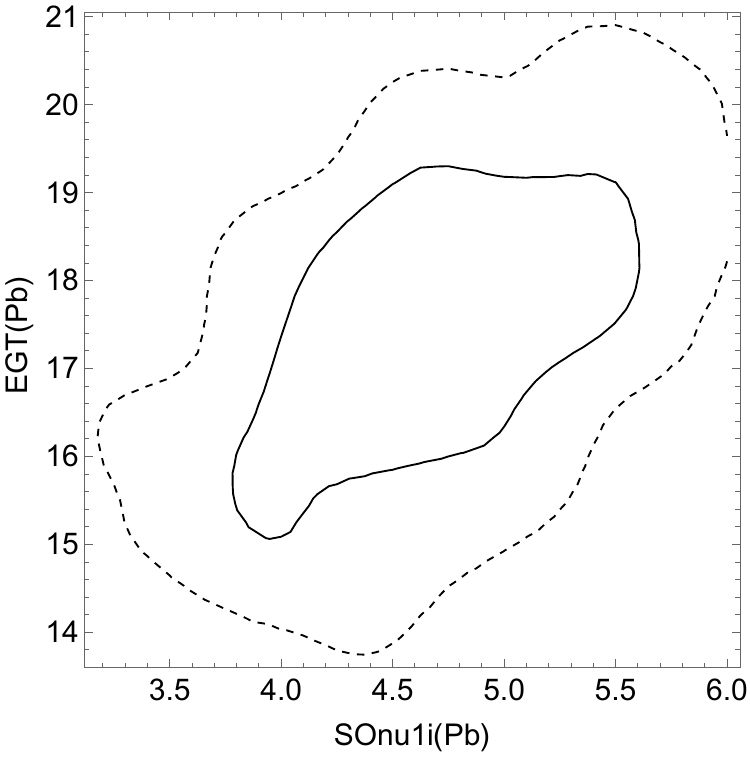}
    \includegraphics[width=0.35\textwidth]{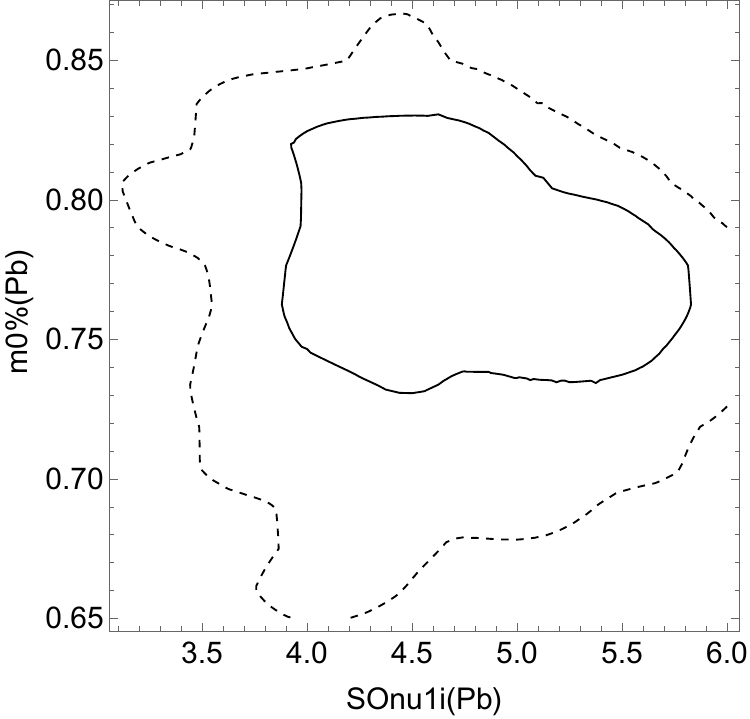}
    \includegraphics[width=0.35\textwidth]{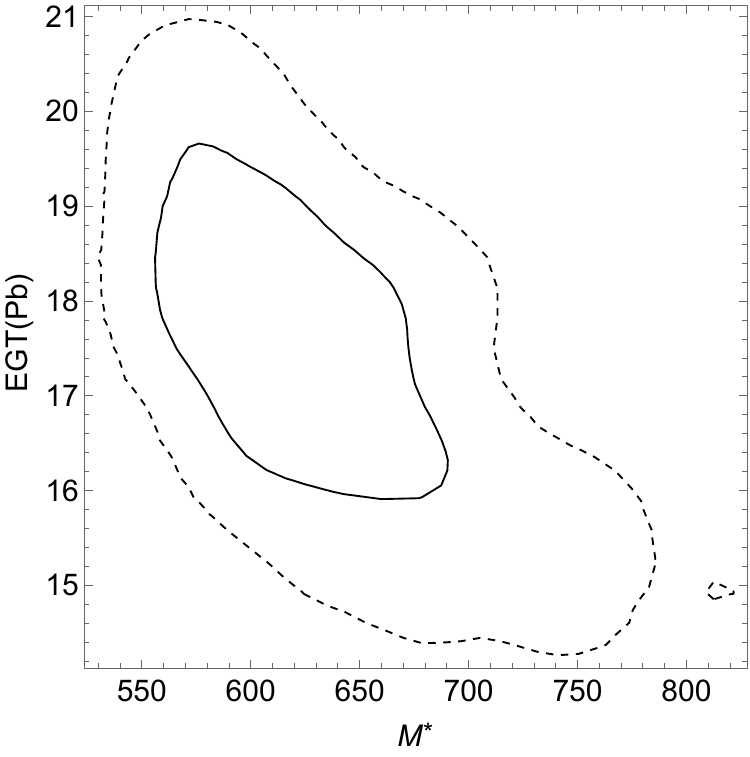}
    \includegraphics[width=0.35\textwidth]{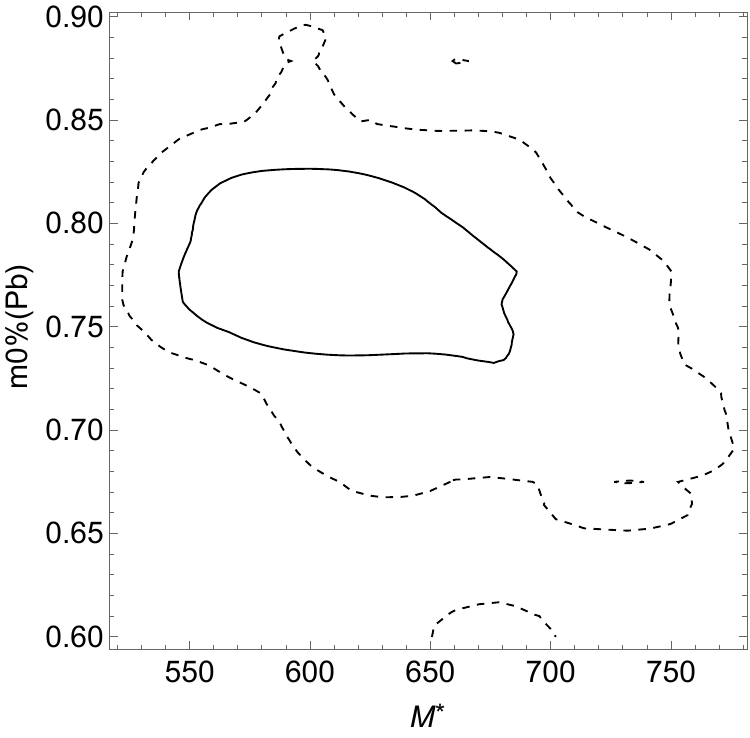}
	\caption{The correlations between GT spectra characteristics and nuclear matter/nuclei properties. The left (right) panel of the first line gives the correlation between $E_\mathrm{GT}$ ($m_0\%$) and the spin orbit splitting $\nu 1i^{-1}_{13/2}$-$\nu 1i_{11/2}$. The left (right) panel of the second line gives the correlation between $E_\mathrm{GT}$ ($m_0\%$) vs $m^*$. }
\label{fig:correlationPhysics}	
\end{figure*}

In Fig.\ref{fig:correlationPhysics}, we present a few correlations that smear the $G'_0~\mathrm{Vs}~E_\mathrm{GT}$ relationship. This suggests that $m_0\%$ is less affected by uncertain physical quantities such as the effective mass and the spin-orbit splitting, compared to $E_\mathrm{GT}$. For a full correlation analysis, the readers are referred to \cite{Zlin2025companion}.

\subsection{Comparison with the extraction of $G_0^\prime$ based on phenomenological models}

To test the robustness of the extraction methodology suggested by traditional phenomenological models, we choose a Skyrme parameterization (SAMi~\cite{Roca-Maza:2012dhp}) as a fiducial model and extracted the value of $G'_0$ following Ref.~\cite{Suzuki:1982usa}. In ~\cite{Suzuki:1982usa}, $G'_0$ is obtained from the GTR energy in the mother nucleus as
\begin{equation}\label{eq:simpleGT}
    E_\mathrm{GTR}-E_d=\epsilon_d+\Delta E(l\sigma)+4\kappa_{\sigma\tau}T_0,
\end{equation}
where $\epsilon_d$ is the average unperturbed excitation energy, $\Delta E(l\sigma)$ is the average spin-orbit splitting value, and $T_0=(N-Z)/2$. The $\kappa_{\sigma\tau}$ is proportional to the Landau-Migdal parameter $G'_0$:
\begin{equation}\label{eq:suzuki}
   A \kappa_{\sigma\tau}=G'_0\frac{k^2_\mathrm{F}}{3m^*}\gamma, 
\end{equation}
where $A$ is the total nucleon number, $k_F=1.36 ~\mathrm{fm}^{-1}$, $m^*$ is the effective mass and $\gamma$ is roughly approximated as $0.5$ in ~\cite{Suzuki:1982usa}. 

\begin{figure*}
	\centering
	\includegraphics[width=0.435\textwidth]{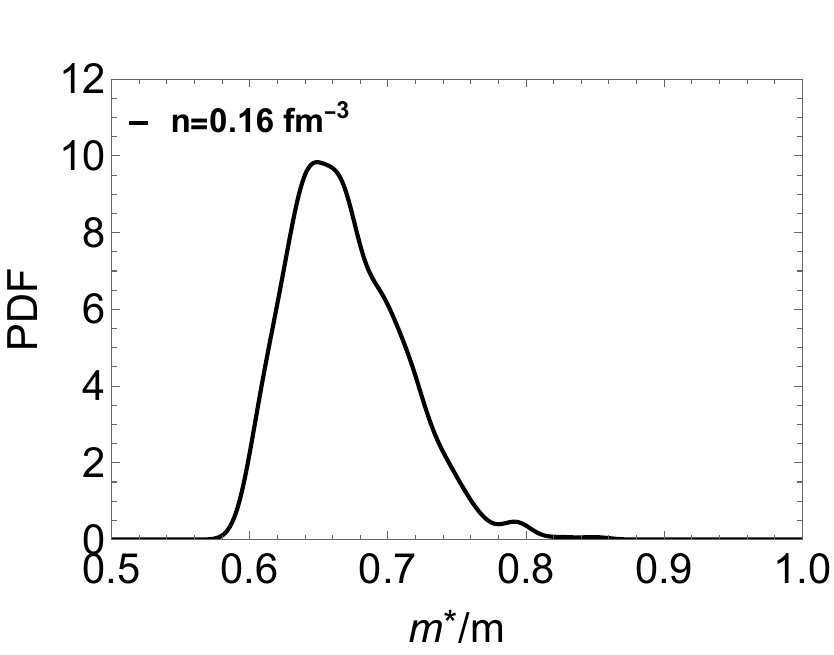}
	\caption{The probability distribution of $m^*/m$ in symmetric nuclear matter at $\rho=0.16~\mathrm{fm^{-3}}$ (right), inferred by the GTAll calculations described in the text. The mean of $m^*/m$ at saturation density is 0.67.
    }
\label{fig:mstar}	
\end{figure*}

We use the prediction of $E_\mathrm{GTR}$, $\epsilon_d$, and $\Delta E(l\sigma)$ by SAMi to deduce $G'_0$ following Eqs. 
(\ref{eq:simpleGT}) and 
(\ref{eq:suzuki}). In this fiducial test, the single-particle states, the GTR spectra and the effective nucleon mass are all self-consistently described by the SAMi Skyrme model. Consequently, the extracted $G'_0$ should be close to the one corresponding to the SAMi force if the method applied in \cite{Suzuki:1982usa} is accurate. For $^{208}\mathrm{Pb}$, SAMi predicts that $E_\mathrm{GTS}-E_\pi=19.2~\mathrm{MeV}$ (which corresponds to $E_\mathrm{GTS}-E_d=15.54~\mathrm{MeV}$),
$\epsilon_\mathrm{d}=7.52~\mathrm{MeV}$, $\Delta E (l\sigma)=2.47 ~\mathrm{MeV}$ and $m^*/m=0.66$. By using the aforementioned SAMi-predicted values as input, we get $G'_0\approx 0.69$. However, the exact $G'_0$ corresponding to the SAMi interaction is 0.35, which is about one-half of the extracted value based on this method. It suggests that the larger $G'_0$ extracted from traditional empirical models may result from the fact that these models used simplified assumptions (e.g., in the case of~\cite{Suzuki:1982usa}, a schematic solution of RPA, constant $G'_0$ and attenuation factor $\gamma$). The main reason for this deviation could be the neglect of the density dependence of $G'_0$ in the schematic models, as discussed in the main manuscript.

Furthermore, empirical models do not self-consistently calculate the effective nucleon mass, and an approximation of $0.8\lessapprox m^*/m\lessapprox 1.0 $ is usually used. 
However, as shown in Fig. \ref{fig:mstar}, the effective masses sampled from GTAll Bayesian inference (where the mean of $m^*/m$ is 0.67) indicate that the effective mass may very likely be smaller than the former assumptions, which induces further disagreement of the extracted $G'_0$ based on self-consistent Skyrme methods and traditional empirical methods. 
\clearpage
\bibliography{References}
\bibliographystyle{apsrev}